\documentstyle[12pt,epsf,psfig]{article}
\renewcommand{\baselinestretch}{1.3}

\begin{document}
\def\be{\begin{eqnarray}}
\def\en{\end{eqnarray}}
\def\non{\nonumber}
\def\la{\langle}
\def\ra{\rangle}
\def\ep{\varepsilon}
\def\ms{{\overline{\rm MS}}}
\def\u{\mu_{\rm fact}}
\def\up{\uparrow}
\def\dw{\downarrow}
\def\gg{\Delta\sigma^{\gamma G}}
\def\lsim{ {\ \lower-1.2pt\vbox{\hbox{\rlap{$<$}\lower5pt\vbox{\hbox{$\sim$}
}}}\ } }
\def\gsim{ {\ \lower-1.2pt\vbox{\hbox{\rlap{$>$}\lower5pt\vbox{\hbox{$\sim$}
}}}\ } }
\def\dk{\partial\!\cdot\!K}
\def\pr{{\sl Phys. Rev.}~}
\def\prl{{\sl Phys. Rev. Lett.}~}
\def\pl{{\sl Phys. Lett.}~}
\def\np{{\sl Nucl. Phys.}~}
\def\zp{{\sl Z. Phys.}~}

\font\el=cmbx10 scaled \magstep2

\vskip 1.5 cm

\centerline{\large\bf Remarks on Next-to-Leading Order Analysis of}
\centerline{\large\bf Polarized Deep Inelastic Scattering Data}
\medskip
\bigskip
\medskip
\centerline{\bf Hai-Yang Cheng}
\medskip
\centerline{ Institute of Physics, Academia Sinica}
\centerline{ Taipei, Taiwan 115, Republic of China}
\bigskip
\bigskip
\bigskip
\bigskip
\centerline{\bf Abstract}
\bigskip

  Since the $x$ dependence of the axial-anomaly effect in inclusive
polarized deep
inelastic scattering is fixed, the transformation from the $\ms$
scheme to different factorization schemes are not arbitrary.
If the quark spin distribution is demanded to be anomaly-free so that
it does not evolve with $Q^2$ and hard gluons contribute to the first
moment of $g_1(x)$, then all the moments of coefficient and 
splitting functions are fixed by perturbative QCD for a given 
$\gamma_5$ prescription, contrary to the commonly used 
Adler-Bardeen (AB) or AB-like scheme.
It is urged that, in order to correctly demonstrate the effect of 
factorization scheme dependence, the QCD analysis of polarized structure 
functions in next-to-leading order should be performed, besides the $\ms$
scheme, in the chiral-invariant factorization scheme in which the axial
anomaly resides in the gluon coefficient function, instead of the less
consistent and ambiguous AB scheme.

\newpage

    Because of the availability of the two-loop polarized splitting 
functions $\Delta P^{(1)}_{ij}(x)$ recently \cite{Mertig}, it became 
possible to embark on a full next-to-leading order (NLO) analysis of the 
experimental data of polarized structure functions by taking into account 
the measured $x$ dependence of $Q^2$ at each $x$ bin. The NLO analyses
have been performed in the $\overline{\rm MS}$ scheme and the
Adler-Bardeen (AB) scheme [2-10]. Of course, physical quantities such as the 
polarized
structure function $g_1(x)$ are independent of choice of the
factorization convention. Physically, the spin-dependent valence quark and 
gluon distributions should be
the same in both factorization schemes. The recent analysis by the E154
Collaboration \cite{E154} has determined the first moments of parton 
spin densities in both schemes:
\be
(\Delta u_v)_\ms=0.69^{+0.03+0.05+0.14}_{-0.02-0.04-0.01}\,, \qquad && (\Delta 
u_v)_{\rm AB}=0.74^{+0.03+0.03+0.07}_{-0.02-0.03-0.01} \, , \non \\
(\Delta d_v)_\ms=-0.40^{+0.03+0.03+0.07}_{-0.04-0.03-0.00} \,, \qquad &&  
(\Delta d_v)_{\rm AB}=-0.33^{+0.02+0.01+0.01}_{-0.04-0.05-0.03}\, , \non \\
\Delta G_\ms=1.8^{+0.6+0.4+0.1}_{-0.7-0.5-0.6}\, , \qquad &&  \Delta 
G_{\rm AB}=0.4^{+1.0+0.9+1.1}_{-0.7-0.6-0.1}\,,  \non\\
\Delta\Sigma_\ms=0.20^{+0.05+0.04+0.01}_{-0.06-0.05-0.01}\, ,   \qquad &&
\Delta\Sigma_{\rm AB}=0.25^{+0.07+0.05+0.05}_{-0.07-0.05-0.02}\, ,
\en
where errors are statistic, systematical and theoretical, and $\Delta\Sigma=
\Delta u+\Delta d+\Delta s$ with 
\be
\Delta q=\Delta q_v+\Delta q_s=\int^1_0dx\Delta q(x)=\int^1_0dx [q^\up(x)+
\bar q^\up(x)-q^\dw(x)-\bar q^\dw(x)].
\en
We see that although the results of the fits in both $\ms$ and AB schemes are
consistent within errors, the central values for $\Delta q_v$, especially 
for $\Delta G$, are
not the same and the fits are significantly less stable in the AB scheme.
This sounds somewhat annoying since if the polarized structure function
is truly factorization scheme independent, then it is expected that the
central values and errors of $\Delta q_v$ and $\Delta G$ in the 
$\overline{\rm MS}$ and AB prescriptions should be quite similar and
that $\Delta\Sigma$ obeys the relation
\be
\Delta\Sigma_\ms=\,\Delta\Sigma_{\rm AB}-{3\alpha_s\over 2\pi}\Delta G_{\rm
AB}.
\en
Also, because the NLO spin-dependent splitting
functions $\Delta P_{ij}^{(1)}(x)$ are originally calculated in
the $\overline{\rm MS}$ scheme, one may wonder if the polarized splitting
kernels in NLO proposed in the AB scheme will render the evolution of
$g_1^p(x,Q^2)$ scheme independent.

   In this short Letter, we wish to emphasize that since the $x$ 
dependence of the axial-anomaly effect in the quark spin distribution
or in the gluon coefficient function, depending on the chosen
factorization scheme, is fixed, the transformation of coefficient
and splitting functions from the $\overline{\rm MS}$
scheme to the improved parton model scheme in which the 
spin-dependent quark distribution does not evolve with $Q^2$ and
hard gluons make contributions to the first moment of $g_1^p(x)$,
is also determined. As a consequence, we urge that a NLO
analysis of $g_1(x,Q^2)$ data should be performed in the 
so-called chiral-invariant factorization scheme to be introduced 
below in order to see if $g_1(x,Q^2),~\Delta
q_v(x,Q^2),~\Delta G(x,Q^2)$ are really scheme independent. Although
none of the material presented in this Letter is new, 
a clarification on this issue is fundamentally important for the QCD
analysis of polarized structure functions in NLO.

   Including QCD corrections to NLO, the generic expression for the polarized 
proton structure function has the form
\be
g_1^p(x,Q^2) &=& {1\over 2}\sum e_q^2\Big[ \Delta C_q(x,\alpha_s)\otimes
\Delta q(x,Q^2)+\Delta C_G(x,\alpha_s)\otimes\Delta G(x,Q^2)\Big] \non\\
&=& {1\over 2}\sum e_q^2\Big[ \Delta q^{(0)}(x,Q^2)+\Delta 
q^{(1)}(x,Q^2)+\Delta q_s^G(x,Q^2)   \\    
&& +\Delta C_q^{(1)}(x,\alpha_s)\otimes\Delta q^{(0)}(x,Q^2)+
\Delta C_G^{(1)}(x,\alpha_s)\otimes\Delta G(x,Q^2)+\cdots\Big],  
\non
\en
where uses of $\Delta C_q^{(0)}(x)=\delta(1-x)$ and $\Delta 
C_G^{(0)}(x)=0$ have been made,  
$\otimes$ denotes the convolution
\be
f(x)\otimes g(x)=\int^1_x {dy\over y}\,f\left({x\over y}\right) g(y),
\en
and $\Delta C_q$, $\Delta C_G$ are short-distance quark and gluon 
coefficient functions, respectively. More specifically, $\Delta C^{(1)}_G$
arises from the hard part of the polarized photon-gluon cross section, 
while $\Delta C^{(1)}_q$ from the short-distance part of the photon-quark
cross section. Contrary to the coefficient functions, $\Delta q_s^G(x)$
and $\Delta q^{(1)}(x)$ come from the soft part of polarized
photon-gluon and photon-quark scatterings. Explicitly, they
are given by 
\be
\Delta q^{(1)}(x,Q^2)=\Delta \phi^{(1)}_{q/q}(x)\otimes \Delta q^{(0)}(x,Q^2), 
\qquad \Delta q_s^G(x,Q^2)=\Delta\phi_{q/G}^{(1)}(x)\otimes\Delta G(x,Q^2),
\en
where $\Delta \phi_{j/i}(x)$ is the polarized distribution of parton $j$
in parton $i$. Diagrammatically, $\Delta\phi_{q/q}^{(1)}$ and
$\Delta\phi_{q/G}^{(1)}$ are depicted in Fig.~1 (see e.g. \cite{Qiu}).
\begin{figure}[h]
\vspace{-3cm}
\hskip 1cm
  \psfig{figure=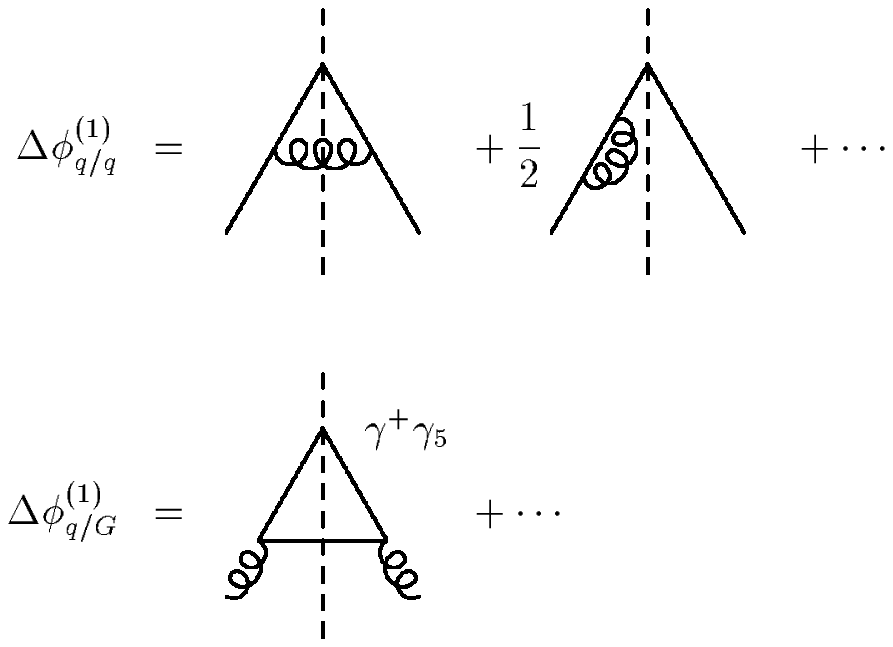,width=13cm}
\vspace{-11cm}
    \caption[]{\small Diagrams for the quark spin distributions inside the
parton: $\Delta\phi^{(1)}_{q/q}$ and $\Delta\phi^{(1)}_{q/G}$.}
    \label{fig1} 

\end{figure}

Since $\Delta\phi^{(1)}$ is ultravioletly divergent, it is clear that, 
just like the case of unpolarized deep inelastic scattering (DIS), the
coefficient functions $\Delta C_q$ and $\Delta C_G$ depend on how 
the parton spin distributions $\Delta \phi^{(1)}_{j/i}$ are defined, 
or how the ultraviolet regulator is specified on $\Delta\phi^{(1)}$.
That is, the ambiguities in defining $\Delta\phi^{(1)}_{q/q}$ and
$\Delta\phi^{(1)}_{q/G}$ are reflected on the ambiguities in extracting
$\Delta C^{(1)}_q$ and $\Delta C^{(1)}_G$. Consequently, the decomposition of
the photon-gluon and photon-quark cross sections into the hard
and soft parts depends on the choice of the factorization scheme and 
the factorization scale $\mu$ [for simplicity, we have set $\mu^2=Q^2$ 
in Eq.~(4)]. Of course, the physical
quantity $g^p_1(x)$ is independent of the factorization
prescription (for a review on the issue of factorization, see 
\cite{Cheng96}).

However, the situation for
the polarized DIS case is more complicated: In addition to all
the ambiguities that spin-averaged parton distributions have,
the parton spin densities are subject to two extra
ambiguities, namely, the axial anomaly and the definition of $\gamma_5$
in $n$ dimension \cite{Qiu}. It is well known that the polarized triangle 
diagram for $\Delta\phi^{(1)}_{q/G}$ (see Fig.~1) has 
an axial anomaly. There are two extreme ultraviolet regulators 
of interest. One of them, which we refer to as the chiral-invariant 
(CI) factorization scheme,
respects chiral symmetry and gauge invariance but not the axial
anomaly. This corresponds to a direct brute-force cutoff $\sim
\mu$ on the $k_\perp$ integration in the triangle diagram ( i.e. 
$k^2_\perp\lsim \mu^2$) with
$k_\perp$ being the quark transverse momentum perpendicular to the
virtual photon direction. Since the gluonic anomaly is manifested at
$k_\perp^2\to\infty$, it is evident that the spin-dependent quark
distribution [i.e. $\Delta q^{(1)}(x)$] in the CI factorization 
scheme is anomaly-free. Note that this is the
$k_\perp$-factorization scheme employed in the usual improved parton model
\cite{Efremov}.

   The other ultraviolet cutoff on the triangle diagram of Fig.~1, as 
employed in the approach of the operator product expansion (OPE), is chosen 
to satisfy gauge symmetry and the gluonic anomaly. As a result, chiral
symmetry is broken in this gauge-invariant (GI) factorization
scheme and a sea polarization is perturbatively induced from hard
gluons via the anomaly. This perturbative mechanism for sea quark
polarization is independent of the light quark masses.
A straightforward calculation gives \cite{Bass,Cheng96}
\be
\Delta\phi_{q/G}^{(1)}(x)_{\rm GI}=\Delta\phi_{q/G}^{(1)}(x)_{\rm CI}
-{\alpha_s\over \pi}(1-x),   \label{phi}
\en
where the term ${\alpha_s\over\pi}(1-x)$ originates from the QCD anomaly
arising from the region where $k_\perp^2\to\infty$. 
Two remarks are in order. First, this term was
erroneously claimed in some early literature \cite{Mank,Ball}
to be a soft term coming from $k_\perp^2\sim m_q^2$ (see below). Second, 
although the quark spin distribution inside the gluon
$\Delta\phi^{(1)}_{q/G}(x)$ cannot be reliably calculated by perturbative
QCD, its difference in GI and CI schemes is trustworthy in QCD.
Since the polarized valence quark distributions are $k_\perp$-factorization
scheme independent, the total quark spin distributions in GI and 
CI schemes are related via Eq.~(6) \cite{Cheng97}
\be
\Delta q(x,Q^2)_{\rm GI}=\Delta q(x,Q^2)_{\rm CI}-{\alpha_s(Q^2)\over 
\pi}\,(1-x)\otimes\Delta G(x,Q^2).    \label{deltaq}
\en
For a derivation of this important result based on a different approach,
namely, the nonlocal light-ray operator technique, see
M\"uller and Teryaev \cite{Muller}. The $x$ dependence of the anomaly 
effect is thus fixed.

 The axial anomaly in the box diagram for polarized photon-gluon
scattering also occurs at $k_\perp^2\to\infty$, more precisely,
at $k^2_\perp=[(1-x)/4x]Q^2$ with $x\to 0$. It is natural to expect that
the axial anomaly resides in the gluon coefficient function $\Delta 
C^{(1)}_G$ in
the CI scheme, whereas its effect in the GI scheme is shifted
to the quark spin density. Since $\Delta C^{(1)}_G(x)$ is the hard part of
the polarized photon-gluon cross section, 
which is sometimes denoted
by $g_1^G(x)$, the polarized structure function of the gluon
target, we have
\be
\Delta C_G^{(1)}(x)=\,g_1^G(x)-\Delta\phi_{q/G}^{(1)}(x).
\en
It follows from Eqs.~(\ref{phi}) and (9) that
\be
\Delta C^{(1)}_G(x)_{\rm GI}=\Delta C^{(1)}_G(x)_{\rm CI}+{\alpha_s\over \pi}
(1-x).   \label{deltag}
\en
It has been argued that the GI scheme is pathologic and inappropriate
\cite{Ball} based on the observation that a direct calculation of $g_1^G(x)$ 
using the mass regulator, for example, for the infrared divergence gives
\be
g_1^G(x)=\,{\alpha_s\over 2\pi}(2x-1)\left( \ln{Q^2\over m^2}+\ln{1-x\over x}
-1\right)+{\alpha_s\over \pi}(1-x), \label{g1g}
\en
where the last term in Eq.~(\ref{g1g}) is an effect of chiral symmetry 
breaking and it arises from the soft region $k_\perp^2\sim m^2$. By comparing
(\ref{deltag}) with (\ref{g1g}), it appears that $\Delta C_G^{(1)}(x)_{\rm
GI}$, which is ``hard" by definition, contains an unwanted ``soft" term. This 
is actually not the case. Choosing a chiral-invariant cutoff on the 
$k_\perp$-integration, a perturbative QCD evaluation yields 
(see \cite{Cheng96} for a review on the detail of derivation)
\be
\Delta C^{(1)}_G(x)_{\rm CI}=\,{\alpha_s\over 2\pi}\left[(2x-1)(\ln{1-x\over
x}-1)\right].  
\en
The ${\alpha_s\over \pi}(1-x)$ term disappears in $\Delta C^{(1)}_G(x)_{\rm 
CI}$, as it should
be, but it emerges again in the GI scheme due to the axial anomaly 
[see Eq.~(\ref{deltag}) or (\ref{phi})] and this time reappears in the 
hard region. That is, the gluon coefficient $\Delta C_G^{(1)}(x)_{\rm GI}$ 
is genuinely hard.

It is easily seen that the first moments of $\Delta C_G(x)$,
$\sum_q\Delta q(x)$ and $g_1^p(x)$ are
given by
\be
&& \int^1_0 dx\Delta C^{(1)}_G(x)_{\rm GI}=0, \qquad \int^1_0 dx\Delta 
C^{(1)}_G(x)_{\rm CI}=-{\alpha_s\over 2\pi},    \\
&& \Delta\Sigma_{\rm GI}(Q^2)=\,\Delta\Sigma_{\rm CI}(Q^2)-{n_f\alpha_s(Q^2)
\over 2\pi}\Delta G(Q^2),
\en
and 
\be
\Gamma^p_1\equiv\int^1_0 g^p_1(x,Q^2)dx &=& {1\over 2}\sum e_q^2\left(\Delta 
q_{\rm CI}(Q^2)-{\alpha_s(Q^2)\over 2\pi}\Delta G(Q^2)\right)  \non \\
&=& {1\over 2}\sum e_q^2\Delta q_{\rm GI}(Q^2),
\en
where $\Delta G\equiv\int^1_0\Delta G(x)dx$,
and we have neglected contributions to $g_1^p$ from $\Delta\phi^{(1)}_{q/q}$ 
and $\Delta C_q^{(1)}$. Note that $\Delta\Sigma_{\rm GI}(Q^2)$ is 
equivalent to the singlet axial charge $\la p,s|J_\mu^5 |p,s\ra$.
The well-known results (12-14) indicate 
that $\Gamma_1^p$ receives anomalous 
gluon contributions in the CI factorization scheme (e.g. the 
improved parton model), whereas hard gluons do not play any role 
in $\Gamma_1^p$ in the GI scheme such as the OPE approach. From
(14) it is evident that the sea quark or anomalous gluon 
interpretation for the suppression of $\Gamma_1^p$ observed experimentally
is simply a matter of convention
\cite{BQ}. From Eqs.~(\ref{deltaq}) and (\ref{deltag}) we see that
$\Delta q^G_s(x)+\Delta C_G^{(1)}(x)\otimes\Delta G(x)$ and hence 
$g_1^p(x)$ is
independent of the choice of the $k_\perp$-factorization scheme, 
as it should be. We would like to stress that physically, the GI and CI
$k_\perp$-factorization schemes are exactly on the same footing, though
philosophically one may argue that it is more natural to have $\Delta C_G$
include all short-distance contributions.

   The $\overline{\rm MS}$ scheme is the most common one chosen in the
GI factorization convention. However, the quark coefficient function 
$\Delta C_q^{(1)}(x)$
in the dimensional regularization scheme is subject to another ambiguity, 
namely, the definition
of $\gamma_5$ in $n$ dimension used to specify the ultraviolet cutoff
on $\Delta\phi_{q/q}^{(1)}$ (see Fig.~1). For example, $\Delta C_q^{(1)}(x)$
calculated in the $\gamma_5$ prescription of 't Hooft and Veltman,
Breitenlohner and Maison (HVBM) is different from that computed 
in the dimension reduction scheme \cite{Qiu}. The result
\be
\Delta C_q^{(1)}(x)=C_q^{(1)}(x)-{2\alpha_s\over 3}(1+x)
\en
usually seen in the literature is obtained in the HVBM scheme, where $C_q(x)$
is the unpolarized quark coefficient function. Of course, the quantity
$\Delta q^{(1)}(x)+\Delta C_q(x)\otimes\Delta q^{(0)}(x)$ and hence 
$g_1^p(x)$ is independent of the definition of $\gamma_5$ in 
dimensional regularization.

   In order to determine the $Q^2$ evolution of the polarized structure
function $g_1(x,Q^2)$ to NLO, it is necessary to know the two-loop
splitting functions $\Delta P_{ij}^{(1)}(x)$ in the NLO evolution
equation. Since the complete results for $\Delta P_{ij}^{(1)}(x)$
are available only in the $\overline{\rm MS}$ scheme, a GI factorization
scheme, it is natural to ask how do we analyze the DIS data of $g_1$
in the CI scheme ? One possibility is proposed in \cite{Cheng97} that
the evolution of the parton spin distributions $\Delta q(x,Q^2)_{\rm
GI}$ and $\Delta G(x,Q^2)$ is first determined from the 
Dokshitzer-Gribov-Lipatov-Altarelli-Parisi (DGLAP) evolution
equation, and then $\Delta q(x,Q^2)_{\rm CI}$ in the CI scheme is related to
$\Delta q(x,Q^2)_{\rm GI}$ and $\Delta G(x,Q^2)$ via Eq.~(\ref{deltaq}). 
The other equivalent possibility, as advocated by M\"uller and
Teryaev \cite{Muller}, is to make a renormalization group 
transformation of NLO splitting kernels and coefficient functions
from the GI scheme to the CI one. In addition to the
coefficient functions [cf. Eq.~(10)]
\be
\Delta C_q^{(1)}(x)_{\rm CI}=\Delta C_q^{(1)}(x)_{\rm GI}, \qquad
\Delta C_G^{(1)}(x)_{\rm CI}=\Delta C_G^{(1)}(x)_{\rm GI}-A(x), \label{cf}
\en
where $A(x)\equiv{\alpha_s\over \pi}(1-x)$, the NLO splitting functions
in the CI scheme can be obtained by applying Eq.~(\ref{deltaq}) to 
the spin-dependent evolution equations:
\be
&& {d\over dt}\Delta q_{\rm NS}(x,t)=\,{\alpha_s(t)\over 2\pi}\Delta P_{qq}^
{\rm NS}(x)\otimes\Delta q_{\rm NS}(x,t),   \non \\
&& {d\over dt}\left(\matrix{\Delta q_{\rm S}(x,t)   \cr   \Delta G(x,t) \cr}
\right)=\,{\alpha_s(t)\over 2\pi}\left(\matrix{\Delta P_{qq}^{\rm S}(x) & 
2n_f\Delta P_{qG}(x)  \cr  \Delta P_{Gq}(x) & \Delta P_{GG}(x) \cr} 
\right)\otimes\left(\matrix
{\Delta q_{\rm S}(x,t)   \cr   \Delta G(x,t) \cr} \right),
\en
where $t=\ln(Q^2/\Lambda^2_{_{\rm QCD}})$, and the indices
S and NS denote singlet and non-singlet parton distributions,
respectively. It is straightforward to show that
\be
\Delta P_{qq}^{\rm NS}(x)_{\rm CI} &=& \Delta P_{qq}^{\rm NS}(x)_{\rm GI},
\qquad \Delta P_{Gq}(x)_{\rm CI}=\Delta P_{Gq}(x)_{\rm GI}, \non \\
\Delta P_{qq}^{\rm S}(x)_{\rm CI} &=& \Delta P_{qq}^{\rm S}(x)_{\rm GI}
+A(x)\otimes \Delta P_{Gq}(x)_{\rm GI},   \non \\
\Delta P_{GG}(x)_{\rm CI} &=& \Delta P_{GG}(x)_{\rm GI}
-A(x)\otimes \Delta P_{Gq}(x)_{\rm GI},   \non \\
2n_f\Delta P_{qG}(x)_{\rm CI} &=& 2n_f\Delta P_{qG}(x)_{\rm GI}
-{\alpha_s\beta\over 4\pi}\,A(x)  \non \\
&& -A(x)\otimes [\Delta P_{qq}^{\rm S}
-\Delta P_{GG}+\Delta P_{Gq}\otimes A](x)_{_{\rm GI}},   \label{kernel}
\en
where
\be
\beta=\left(11-{2\over 3}n_f\right)+{\alpha_s\over 2\pi}\left(51-
{19\over 3}n_f\right)+\cdots
\en
is the usual $\beta$-function.
The above results are first obtained by M\"uller and Teryaev \cite{Muller}.
In short, 
Eqs.~({\ref{cf}) and (\ref{kernel}) provide the NLO coefficient 
and splitting functions necessary for the CI factorization scheme. It is
easy to check that $\Delta q_{\rm CI}\equiv\int^1_0dx\Delta q(x)_{\rm
CI}$ does not evolve with $Q^2$ and that hard gluons contribute to
$\Gamma_1^p$ in an amount as shown in Eq.~(14).

  Several CI-like schemes were proposed in \cite{Ball} in which 
the singlet anomalous dimension and the first moments of the gluon 
coefficient function are fixed to be
\be
\int^1_0dx\Delta C_G^{(1)}(x)=-{\alpha_s\over 2\pi}, \qquad
\Delta \gamma_{{\rm S},qq}^{(1),1}=0,   
\en
where 
\be
\Delta\gamma_{ij}^n=\int^1_0\Delta P_{ij}(x)x^{n-1}dx=\Delta\gamma_{ij}^{(0),
n}+{\alpha_s\over 2\pi}\Delta\gamma_{ij}^{(1),n}+\cdots.
\en
The remaining moments of the coefficient functions and anomalous dimensions
are then constructed by modifying the counterparts in the 
$\overline{\rm MS}$ scheme. Three of such schemes, namely the
Adler-Bardeen (AB) scheme, the off-shell scheme, and the 
Altarelli-Ross scheme were considered in \cite{Ball}. For
example, higher moments in the AB scheme are fixed by
requiring that the full scheme change from the $\ms$ scheme to the 
AB scheme be independent of $x$, so that the large and small $x$
behavior of the coefficient functions is unchanged. Evidently, this
is in contradiction to the $(1-x)$ behavior of the anomaly effect
shown in Eq.~(\ref{cf}); that is, though the first moments of $\Delta q(x)$
and $g_1(x)$ in the AB scheme are in agreement with the CI 
factorization scheme, this is no longer true for other moments. In
principle, a measurement of higher moments of spin-dependent parton 
distributions will enable us to discern between CI and AB schemes. Since
the AB scheme is not consistent with perturbative QCD for the $x$
dependence of the renormalization group transformation for coefficient
and splitting functions, it is thus not pertinent to use this factorization 
convention to analyze the polarized structure functions in NLO.

   To conclude, since the $x$ dependence of the axial-anomaly effect 
is fixed, the transformation of spin-dependent coefficient and splitting
functions from the $\ms$ scheme to the improved parton model 
(or chiral-invariant) factorization scheme in which the axial 
anomaly is shifted to the gluon coefficient function
are uniquely determined by perturbative QCD for a given $\gamma_5$ 
prescription, contrary to the commonly used 
Adler-Bardeen scheme. We thus believe that the CI scheme should be used,
instead of the less consistent and ambiguous AB scheme.
In order to see the factorization scheme 
independence of $\Delta q_v(x,Q^2),~\Delta G(x,Q^2)$ and $g_1(x,Q^2)$ and
demonstrate the effect of scheme dependence for $\Delta q_s(x,Q^2),~
\Delta\Sigma(x,Q^2)$, it is urged that
the QCD analysis of polarized structure 
functions in NLO should be carried out in both $\ms$ (or any gauge-invariant)
and chiral-invariant factorization schemes. The $Q^2$ evolution
of parton spin distributions in the latter can be obtained by either 
studying the $Q^2$ evolution first in the $\ms$ scheme and then applying
Eq.~(\ref{deltaq}) afterwards or solving the DGLAP equation directly 
in the CI scheme 
using the splitting functions given by (\ref{kernel}).

\vskip 1.5 cm
\noindent ACKNOWLEDGMENT:~~This work was supported in part by the National 
Science Council of ROC under Contract No. NSC87-2112-M-001-048.

\renewcommand{\baselinestretch}{1.1}
\newcommand{\bi}{\bibitem}
%

\newpage

\end{document}